\documentclass[sn-aps]{sn-jnl}

\usepackage{graphicx}%
\usepackage{multirow}%
\usepackage{amsmath,amssymb,amsfonts}%
\usepackage{amsthm}%
\usepackage{mathrsfs}%
\usepackage[title]{appendix}%
\usepackage{xcolor}%
\usepackage{textcomp}%
\usepackage{manyfoot}%
\usepackage{booktabs}%
\usepackage{algorithm}%
\usepackage{algorithmicx}%
\usepackage{algpseudocode}%
\usepackage{listings}%
\usepackage[compat=1.1.0]{tikz-feynman}
\usepackage{siunitx,adjustbox}
\usepackage{subcaption,geometry,rotating}
\usepackage{slashed,bbold,mathtools,sidecap,tikz,bm,ulem,enumitem,array}
\definecolor{lime}{HTML}{A6CE39}
\DeclareRobustCommand{\orcidicon}{\hspace{-2.1mm}
\begin{tikzpicture}
\draw[lime,fill=lime] (0,0.0) circle [radius=0.13] node[white] {{\fontfamily{qag}\selectfont \tiny ID}}; \draw[white,fill=white] (-0.0525,0.095) circle [radius=0.007]; 
\end{tikzpicture} \hspace{-3.7mm} }
\foreach \x in {A, ..., Z} {\expandafter\xdef\csname orcid\x\endcsname{\noexpand\href{https://orcid.org/\csname orcidauthor\x\endcsname} {\noexpand\orcidicon}}}

\theoremstyle{thmstyleone}%
\theoremstyle{thmstyletwo}%

\theoremstyle{thmstylethree}%

\begin{document}
\title[Article Title]{Growing Evidence for a Higgs Triplet}
\author[1,2]{Andreas Crivellin\orcidD{}}
\email{andreas.crivellin@cern.ch}

\author[3,4]{Saiyad Ashanujjaman\orcidA{}}
\email{s.ashanujjaman@gmail.com}

\author[1,2]{Sumit Banik\orcidB{}}
\email{sumit.banik@psi.ch}

\author[1,2]{Guglielmo Coloretti\orcidC{}}
\email{guglielmo.coloretti@physik.uzh.ch}

\author[5,6]{Siddharth P.~Maharathy\orcidE{}}
\email{siddharth.m@iopb.res.in}

\author[7,8]{Bruce Mellado}
\email{bmellado@mail.cern.ch}


\affil[1]{Physik-Institut, Universität Zürich, Winterthurerstrasse 190, CH–8057 Zürich, Switzerland}

\affil[2]{Paul Scherrer Institut, CH–5232 Villigen PSI, Switzerland}

\affil[3]{Institute of High Energy Physics, Chinese Academy of Sciences, Beijing 100049, China}

\affil[4]{Kaiping Neutrino Research Center, Jiangmen 529399, China}

\affil[5]{Institute of Physics, Bhubaneswar, Sachivalaya Marg, Sainik School, Bhubaneswar 751005, India}

\affil[6]{Homi Bhabha National Institute, Training School Complex, Anushakti Nagar, Mumbai 400094, India}

\affil[7]{School of Physics and Institute for Collider Particle Physics, University of the Witwatersrand, Johannesburg, Wits 2050, South Africa}

\affil[8]{iThemba LABS, National Research Foundation, PO Box 722, Somerset West 7129, South Africa}



\abstract{

With the discovery of a Higgs boson with a mass of 125 gigaelectronvolts (GeV) at the Large Hadron Collider (LHC) at CERN in 2012, the Standard Model (SM) is complete, and despite intensive searches, no new fundamental particle has been observed since then. In fact, a discovery can be challenging without a predictive new physics model because different channels and observables cannot be combined directly and unambiguously. Furthermore, without supporting indirect hints, the signal space to be searched is huge, resulting in diluted significances due to the look-elsewhere effect. 

Several LHC processes with multiple leptons in the final state point towards the existence of a new Higgs boson with a mass between 140\,GeV to 160\,GeV decaying mostly to $W$ bosons. While the former strongly reduces the look-elsewhere effect, the latter indicates that it could be a Higgs triplet with zero hypercharge. Within this simple and predictive extension of the Standard Model, we simulate and combine different channels of di-photon production in association with leptons, missing energy, jets, {\it etc}. Using the full run-2 results by ATLAS, including those presented recently at the Moriond conference, an increased significance of 4.3 standard deviations is obtained for a $\approx$152\,GeV Higgs. Due to the previously predicted mass range, the look-elsewhere effect is negligible, and this constitutes the highest statistical evidence for a new narrow resonance obtained at the LHC. Furthermore, the model predicts a heavier-than-expected $W$ boson, as indicated by the global electroweak fit. If further substantiated, the discovery of a new Higgs would overthrow the SM, provide a compelling case for the construction of future particle colliders, and open the way to a novel understanding of the known shortcomings of the SM. In particular, the triplet Higgs field can lead to a strong first-order phase transition and could thus be related to the matter anti-matter asymmetry in our Universe.}

\pacs[Preprint Numbers]{ ZU-TH 24/24, PSI-PR-24-11, ICPP-81}
\maketitle

\section{Introduction}
\label{sec:intro}

The established theoretical description of Nature at microscopic scales, the Standard Model (SM), with its gauge symmetry group $SU(3)_c\times SU(2)_L\times U(1)_Y$ includes the strong, weak and electromagnetic interactions of matter particles (spin-1/2 fermions) via mediators (spin-1 particles, called gauge bosons). The gauge bosons of the electromagnetic and strong interactions---the photon and gluons---are massless, while those of the weak interactions, the $W$ and $Z$ bosons, are heavy.\footnote{While the gluons ($W$ bosons) originate only from the $SU(3)_c$ ($SU(2)_L$) factor, the photon ($\gamma$) and the $Z$ are an admixture of hypercharge ($U(1)_Y$) and $SU(2)_L$.} The SM has been (with a few exceptions~\cite{Crivellin:2023zui}) very successful in describing the results of the vast majority of particle physics experiments~\cite{ParticleDataGroup:2022pth}. 

In 2012, the final missing particle of the SM, the Higgs boson, was observed by the ATLAS~\cite{ATLAS:2012yve} and CMS~\cite{CMS:2012qbp,CMS:2013btf} collaborations at the Large Hadron Collider (LHC) at CERN. It is important to recall that this discovery took place several decades after the Higgs was predicted by the mechanism of spontaneous symmetry breaking via a scalar field, as proposed in 1964 by Brout and Englert~\cite{Englert:1964et}, Higgs~\cite{Higgs:1964ia,Higgs:1964pj} and Guralnik, Hagen and Kibble~\cite{Guralnik:1964eu}, and implemented in the SM by Glashow~\cite{Glashow:1961tr}  Weinberg~\cite{Weinberg:1967tq} and Salam~\cite{Salam:1968rm}. The reason why finding the Higgs was so difficult is that its mass of 125\,GeV lies beyond the reach of electron-positron colliders like LEP~\cite{LEPWorkingGroupforHiggsbosonsearches:2003ing} at CERN, and its suppressed production cross-section prevented an observation at the Tevatron~\cite{CDF:2012laj} at Fermilab. In fact, the Higgs discovery at the LHC was only possible this time because its signal strengths are predicted by the SM, so CMS and ATLAS were able to combine the di-photon ($\gamma\gamma$) with the 4-lepton channel. Furthermore, the obtained Higgs mass was consistent with the expectation from the global electroweak fit, which provided a search range and indirect confirmation that it is really the SM Higgs. This substantiates the importance of an ultraviolet-complete and predictive model as well as indirect hints for new particles to facilitate discovery in statistically limited searches.

On the theoretical side, the Higgs boson of the SM is an elementary scalar, a type of particle that had never been observed before. It gives masses to the $W$ and $Z$ bosons while keeping the theory renormalizable ({\it i.e.}~theoretically consistent) as shown by 't Hooft and Veltman~\cite{tHooft:1971akt,tHooft:1972tcz} in 1971. Furthermore, the elementary fermions acquire their masses via their so-called Yukawa interactions with the Higgs field~\cite{Weinberg:1967tq,Nambu:1961tp}, which is an essential requirement for the existence of complex structures and processes in our Universe.

However, the SM cannot be the ultimate theory of Nature. For example, it does not account for the fact that more gravitationally interacting matter than visible matter is observed at astrophysical scales, leading to the conjecture of the existence of Dark Matter. Non-vanishing neutrino masses necessitated by neutrino oscillations also require an extension of the SM. Furthermore, the dominance of matter over anti-matter in the universe cannot be explained, and the SM does not include gravity. Therefore, it should be considered an effective description which needs to be superseded by a more comprehensive theory. While no unambiguous direction for such an extension has been established, any heavy new physics poses a problem for fundamental scalar particles because they are subject to quantum corrections involving the corresponding scale, which can be many orders of magnitude larger than the electroweak (EW) scale ($\sim 100$\,GeV). Therefore, the mass of the Higgs boson is puzzlingly small. Moreover, no theoretical principle or symmetry requirement guarantees the minimality of the SM Higgs sector. Solving these puzzles is part of the motivation for many new physics models and future experiments and accelerators.

Additional scalar bosons must play a subleading role in the spontaneous breaking of the weak interactions. One reason for this is that in the SM, the $W$ mass can be calculated in terms of the $Z$ mass and the weak and electromagnetic interaction strengths, and the result agrees well (on an absolute scale) with the corresponding measurement. Nonetheless, the $W$ boson was found to be slightly heavier than expected by the CDF-II collaboration at Fermilab, which prefers a small new physics contribution.\footnote{While the tension with the SM prediction is driven by the CDF-II result, the less precise LHC~\cite{LHCb:2021bjt,ATLAS:2024erm} and LEP~\cite{ALEPH:2013dgf} values are in better agreement with the SM and in some conflict with the CDF result. Inflating the error according to the PDG method, a tension of $3.7\sigma$ is found~\cite{deBlas:2022hdk,Athron:2022isz,Bagnaschi:2022whn}.} In this context, the scalar $SU(2)_L$ triplet with hypercharge 0~\cite{Ross:1975fq,Gunion:1989ci,Chankowski:2006hs,Blank:1997qa,Forshaw:2003kh,Chen:2006pb,Chivukula:2007koj,Bandyopadhyay:2020otm} is particularly interesting since it is the most minimal extension of the SM which predicts a positive definite shift in the $W$ mass at tree-level ({\it i.e.}~at leading order in the expansion of quantum corrections)~\cite{Chabab:2018ert,FileviezPerez:2022lxp,Cheng:2022hbo,Chen:2022ocr,Rizzo:2022jti,Chao:2022blc,Wang:2022dte,Shimizu:2023rvi,Lazarides:2022spe,Senjanovic:2022zwy,Crivellin:2023gtf,Chen:2023ins,Ashanujjaman:2023etj}. Furthermore, it has been shown that it can lead to a strong first-order phase transition~\cite{Patel:2012pi,Niemi:2018asa,Niemi:2020hto} which is an essential ingredient of weak-scale Baryogenesis, a mechanism that can explain the matter-antimatter asymmetry in the universe. Last but not least, the so-called ``multi-lepton anomalies''~\cite{Fischer:2021sqw,Crivellin:2023zui} suggest the existence of a new scalar with a mass range of 140\,GeV to 160\,GeV~\cite{vonBuddenbrock:2016rmr,vonBuddenbrock:2017gvy,Buddenbrock:2019tua,Banik:2023vxa} which decays dominantly to $W$ bosons and is produced in association with lepton, bottom quarks and missing energy~\cite{Crivellin:2021ubm}. Both the production and decay modes are in agreement with the $Y=0$ triplet hypothesis~\cite{Ashanujjaman:2024pky}. Importantly, these indirect hints for a new scalar reduce the look-elsewhere effect. 

\begin{figure*}[t!]
\centering
\includegraphics[scale=0.36]{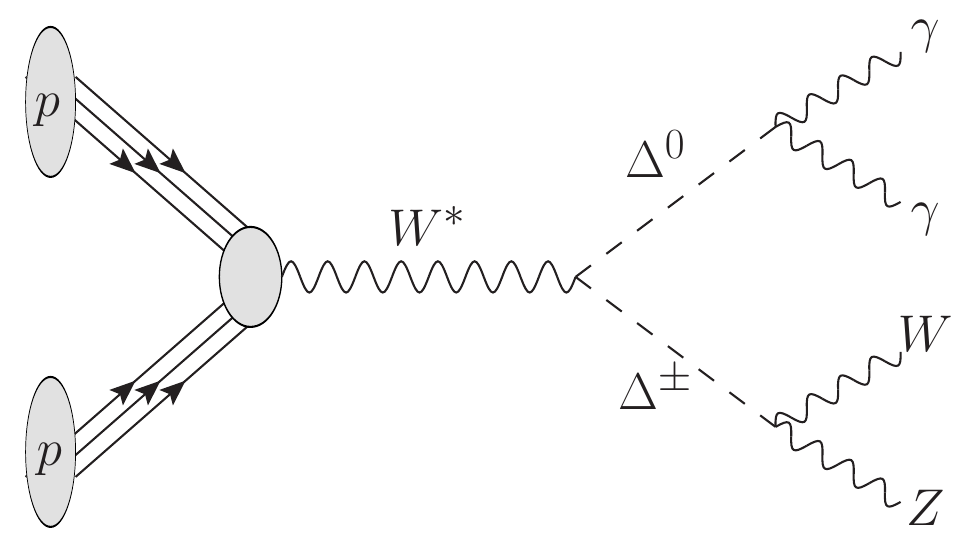}
\hspace{0.24cm}
\includegraphics[scale=0.36]{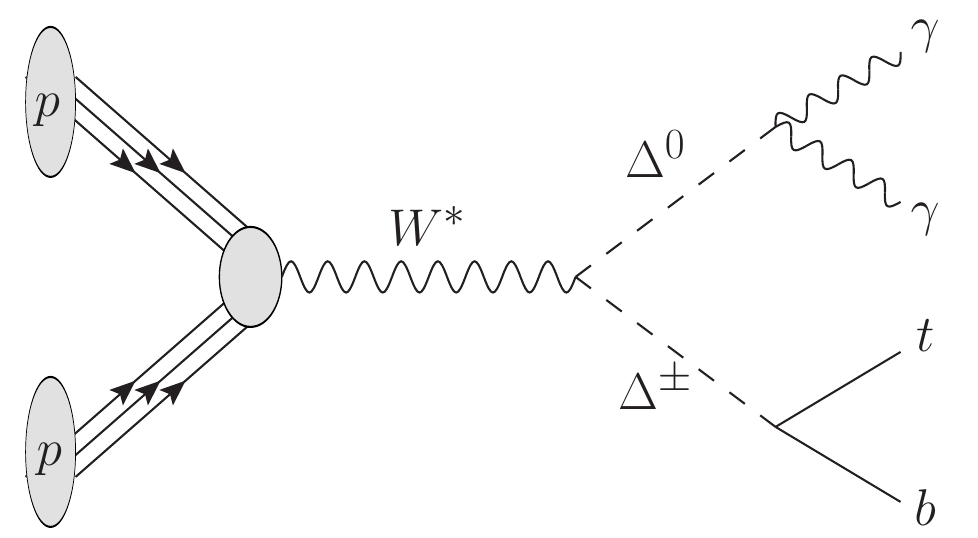}
\caption{Feynman diagrams showing the Drell-Yan production and decays of the triplet Higgses: $pp\to W^*\to (\Delta^{\pm}\to tb,WZ)(\Delta^{0}\to\gamma\gamma)$, which we search for using the side-bands of the SM Higgs analyses of ATLAS~\cite{ATLAS:2023omk,ATLAS-CONF-2024-005}.}
\label{fig:ppToHpmH0_alternate}
\end{figure*}

The SM extended by a $SU(2)_L$ triplet with zero hypercharge is a very predictive model since it contains (in addition to the SM) only one neutral and one charged scalar, without direct couplings to SM fermions. This leads to suppressed production rates such that it can evade LEP and current LHC bounds~\cite{Butterworth:2023rnw}. However, it has distinct collider signatures due to its unavoidable production in proton-proton collisions via off-shell EW gauge bosons and the photon, called Drell-Yan production. This leads to the associated production of di-photons with leptons, missing energy and/or jets~\cite{Ashanujjaman:2024pky} (see Fig.~\ref{fig:ppToHpmH0_alternate}). Note that searching for these exclusive signatures significantly improves the new physics sensitivity by reducing the SM backgrounds due to the requirement of additional particles in the final state. 

In this article, we search for the scalar triplet in associated production channels with di-photons in the mass range suggested by the multi-lepton anomalies. For this, we use the comprehensive ATLAS analysis of Ref.~\cite{ATLAS:2023omk} as well as the latest result for di-photons plus single tau and single lepton channels of Ref.~\cite{ATLAS-CONF-2024-005} presented at the Moriond conference.\footnote{With respect to Ref.~\cite{Ashanujjaman:2024pky}, we will not only include these new data but also perform background refitting and take into account the statistical correlations among the channels.}

\section{Model and Setup}
\label{sec:model}

The SM supplemented by a $SU(2)_L$ triplet scalar with hypercharge 0 is commonly referred to as the $\Delta$SM~\cite{Ross:1975fq,Gunion:1989ci,Chankowski:2006hs,Blank:1997qa,Forshaw:2003kh,Chen:2006pb,Chivukula:2007koj,Bandyopadhyay:2020otm}.\footnote{For details and definitions of the model as well as the calculation of the branching ratios to photons, see Ref.~\cite{Ashanujjaman:2023etj}.} During spontaneous electroweak symmetry breaking, the $SU(2)_L$ doublet SM Higgs and the triplet Higgs acquire their vacuum expectation values $v$ and $v_\Delta$, respectively. While the former gives rise to both $m_W$ and $m_Z$, the latter contributes only to $m_W$; it is thus said to violate the custodial symmetry. More specifically, it leads to a positive definite shift in $m_W$:
\begin{align}
m_W\approx m_W^{\rm SM}\left(1 + \frac{2v_\Delta^2}{v^2}\right),
\end{align}
{\it w.r.t.}~the SM prediction~\cite{ParticleDataGroup:2022pth}. 
This is in agreement with the current global average for the $W$ mass~\cite{ParticleDataGroup:2022pth}, which indicates a positive effect of $\approx$20\,MeV with a significance of $3.7\sigma$~\cite{deBlas:2022hdk}. This implies $v_\Delta \sim \mathcal{O}({\rm GeV})$ such that $v_\Delta\ll v$. Note that the exact value for $v_\Delta$ is immaterial for this work.

The $\Delta$SM contains, in addition to the SM(-like) Higgs ($h$), a charged Higgs ($\Delta^\pm$) and a neutral one ($\Delta^0$). Because $h$ and $\Delta^0$ have the same quantum numbers, they mix after EW symmetry breaking by an angle $\alpha$, {\it i.e.}~the mass eigenstates are linear superpositions of the neutral components of the triplet and doublet Higgses (the interaction eigenstates). However, since this mixing is generally small,\footnote{Measurements of the SM Higgs signal strength, in particular $\gamma\gamma$ and $Z\gamma$, and theoretical constraints such as perturbative unitarity, restrict the mixing angle to be small.} we will use the same labels for the mass and interaction eigenstates. Furthermore, because the mass splitting between $\Delta^0$ and $\Delta^\pm$ is of the order of $v_\Delta$, we can assume both components to be degenerate, {\it i.e.}~$m_{\Delta^0}\approx m_{\Delta^\pm}\equiv m_{\Delta}$, as far as LHC searches are concerned.

\begin{figure*}[t!]
\centering
\includegraphics[scale=0.6]{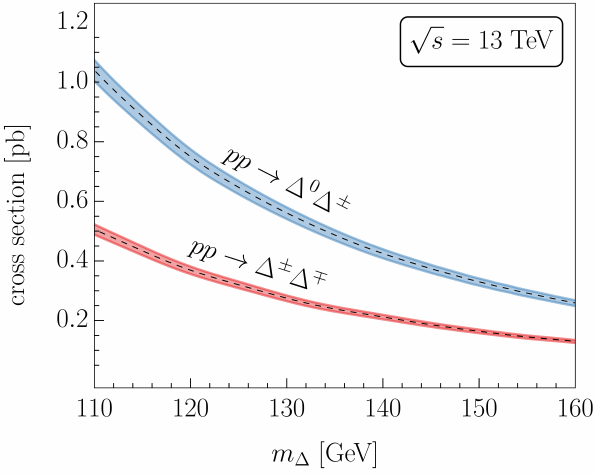}
\hspace{0.3cm}
\includegraphics[scale=0.6]{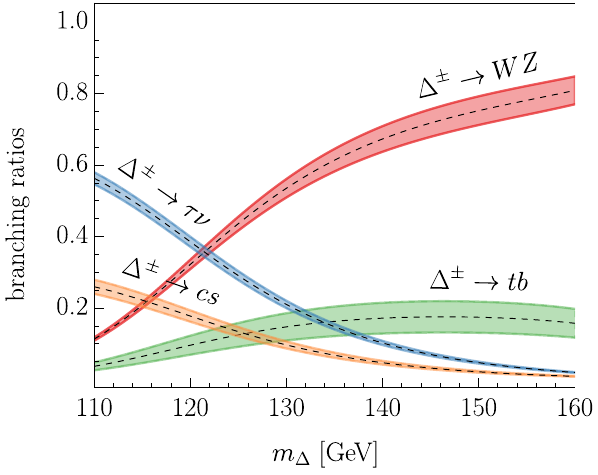}
\caption{Left: Production cross-section for $pp\to \Delta^0 \Delta^\pm$ and $pp\to \Delta^\pm \Delta^\mp$ as a function of the triplet mass including the NNLL and NLO QCD correction factor and uncertainties of Refs.~\cite{Ruiz:2015zca,AH:2023hft}. Right: Dominant branching ratios of the charged component $\Delta^\pm$ as a function of its mass. The errors are estimated from the decays for a SM Higgs with a higher (hypothetical) mass from $h\to tt^*,ZZ^*$ and $h\to cc$~\cite{LHCHiggsCrossSectionWorkingGroup:2013rie}. While these uncertainties are sizable, we checked that their impact on the final significance is very small ($\approx$0.1$\sigma$).}
\label{fig:xsec}
\end{figure*}

Due to its quantum numbers, the triplet Higgs cannot have direct couplings to quarks or leptons. Consequently, it is dominantly produced at the LHC via the Drell-Yan processes $pp\to W^*\to \Delta^0\Delta^\pm$ and $pp\to Z^*/\gamma^*\to \Delta^\pm\Delta^\mp$,\footnote{Note that vector-boson fusion is suppressed by $v_\Delta/v$ and/or $\alpha$, while the latter also suppresses production via gluon fusion.} since it transforms non-trivially under $SU(2)_L$ (see Fig.~\ref{fig:ppToHpmH0_alternate}). Because the couplings of $\Delta^\pm$ to SM particles are mixing-induced by $v_\Delta$, the dominant decay modes are $WZ$, $tb$, $\tau\nu$ and $cs$, and the only free parameter is $m_\Delta$ entering through phase space factors.\footnote{Note that for our mass range of interest, the top and either $Z$ or $W$ are off-shell. However, for convenience, we omitted the asterisk signalling this. In the simulation, we generated both $\Delta^\pm\to W^*Z$ and $\Delta^\pm\to WZ^*$ weighted by their branching ratios.} The resulting branching ratios are shown in Fig.~\ref{fig:xsec}. The dominant decay widths of $\Delta^0$ ($WW$, $bb$ and $ZZ$ for our mas range of interest) depend on $v_\Delta$ and $\alpha$. However, we are interested in $\Delta^0\to\gamma\gamma$ which constitutes a particularly clean signature with controlled backgrounds in experiments. This decay is loop-induced and depends, in addition to $\alpha$ and $v_\Delta$, critically on $m_{\Delta^0}-m_{\Delta^\pm}$ (because the mass difference is related to the trilinear couplings $\Delta^0\Delta^\pm\Delta^\mp$).\footnote{Furthermore, this decay is sensitive to extensions of the $\Delta$SM, which is neither the case for the DY production mechanism nor for the decays of the charged Higgs (which happen at tree-level).} Therefore, in the following, we subsume these parameter dependencies into the di-photon branching ratio Br$(\Delta^0\to\gamma\gamma)$ and consider the latter as a free parameter. 

\begin{table}[t!]
\begin{tabular}{m{1.5cm}m{2.3cm}m{4.8cm}m{1.5cm}}
\hline
\\[-.1cm]
Target & Signal region & Detector level & Correlation\\
\\[-.1cm]
\hline
\\[-.1cm]
\multirow{2}{1.5cm}{High jet activity\cite{ATLAS:2023omk}} & $\ge 4j$ & $n_{\text jet}~\ge 4$ , $|\eta_{\rm jet}|~<2.5$  &$-$\\
&&\\
\\[-.1cm]
\hline
\\[-.1cm]
\multirow{2}{1.5cm}{Top\cite{ATLAS:2023omk}} & $\ell b$ & $n_{\ell=e,\mu} \ge 1$, $n_{b\text{-jet}} \ge 1$ &\multirow{2}{1.5cm}{$-$}\\
& $t_\text{lep}$ &  $n_{\ell=e,\mu}=1$, $n_\text{\rm jet}=n_{b\text{-jet}}=1$ &\\
\\[-.1cm]
\hline
\\[-.1cm]
\multirow{3}{1.5cm}{Lepton} & $2\ell$\cite{ATLAS:2023omk} & $ee,\mu\mu$ or $e\mu$ &\multirow{3}{1.5cm}{$<26\%$}\\
& $1\ell$\cite{ATLAS-CONF-2024-005}& $n_{\ell=e,\mu}=1$, $n_{\tau_{\text{had}}} = 0 $, $n_{b\text{-jet}}=0$,   $E_\text{T}^\text{miss} > 35\text{ GeV}$ (only for $e$-channel) &\\
\\[-.1cm]
\hline
\\[-.1cm]
\multirow{2}{4cm}{$E_\text{T}^\text{miss}$\cite{ATLAS:2023omk}} & $E_\text{T}^\text{miss} > 100 \text{ GeV}$  & $E_\text{T}^\text{miss} > 100 \text{ GeV}$ & \multirow{2}{1.5cm}{$29\%$}\\
& $E_\text{T}^\text{miss} > 200 \text{ GeV}$  & $E_\text{T}^\text{miss} > 200 \text{ GeV}$  &\\
\\[-.1cm]
\hline
\\[-.1cm]
\multirow{1}{4cm}{Tau \cite{ATLAS-CONF-2024-005}} & 1$\tau_{\text{had}}$  & $n_{\ell=e,\mu} = 0 $, $n_{\tau_{\text{had}}} = 1 $, $n_{b\text{-jet}}=0$,  $E_\text{T}^\text{miss} > 35 \text{ GeV}$ &$-$\\
\\[-.1cm]
\hline
\end{tabular}
\caption{The signal regions of the ATLAS analyses~\cite{ATLAS:2023omk,ATLAS-CONF-2024-005} which are sensitive to the Drell-Yan production of the scalar triplet within our mass range of interest. $E_T^{\rm miss}$ stands for missing transverse energy, $t$ for the top quark and $\ell=e,\mu$ for an electron or a muon. $n_{(b)}\text{-\rm jet}$ denotes the number of (bottom quark-initiated) jets and $\eta_{\rm jet}$ the rapidity of the jet. The subscripts `had' and `lep' stand for the corresponding hadronic decays of tau leptons and the leptonic decays of top quarks.}
\label{categories}
\end{table}

\section{Analysis and results}

\begin{figure}[htb!]
\centering
\includegraphics[scale = 0.32]{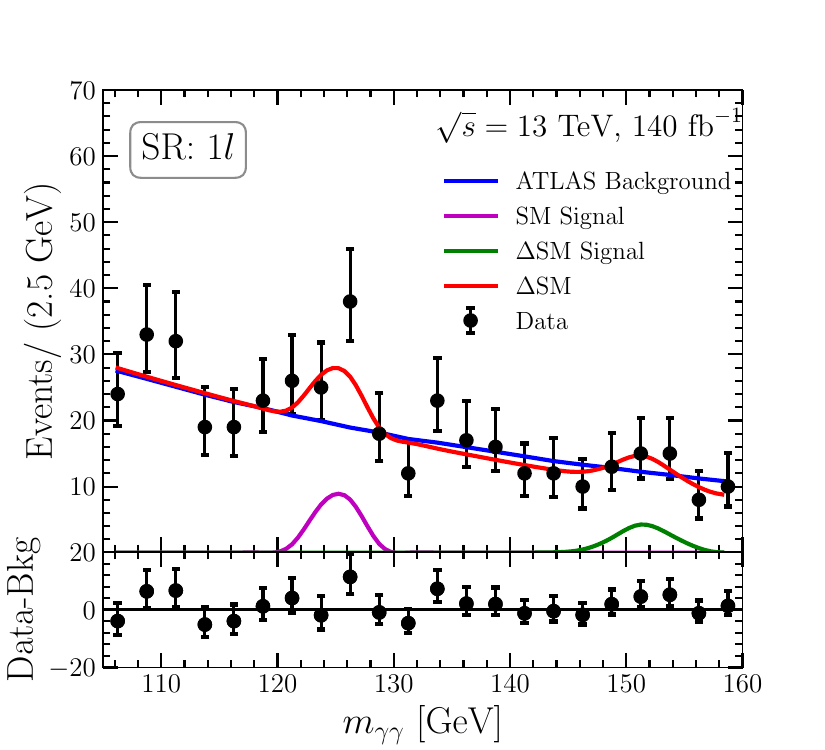}
\includegraphics[scale = 0.32]{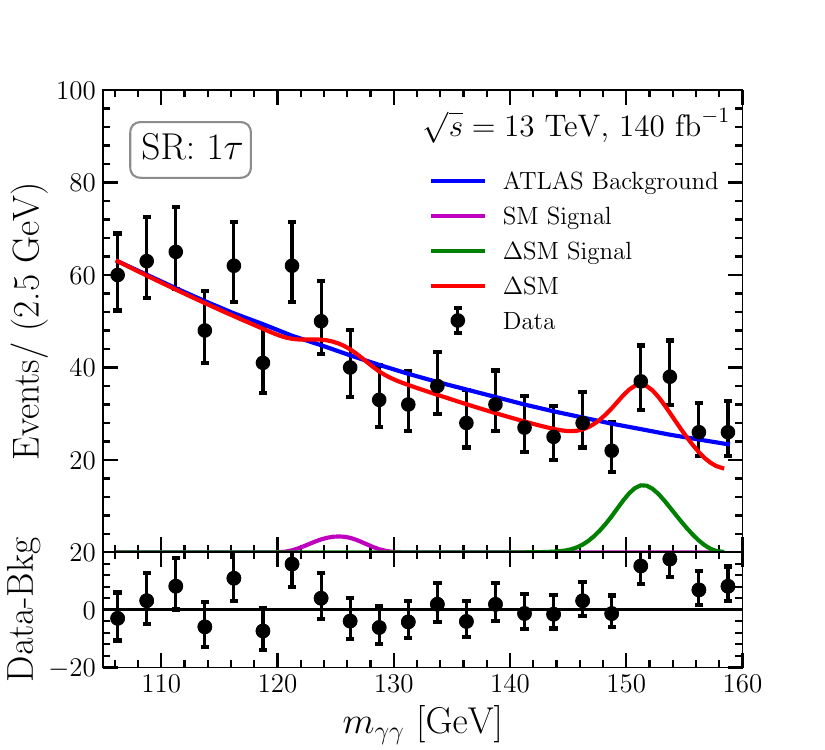}
\includegraphics[scale = 0.32]{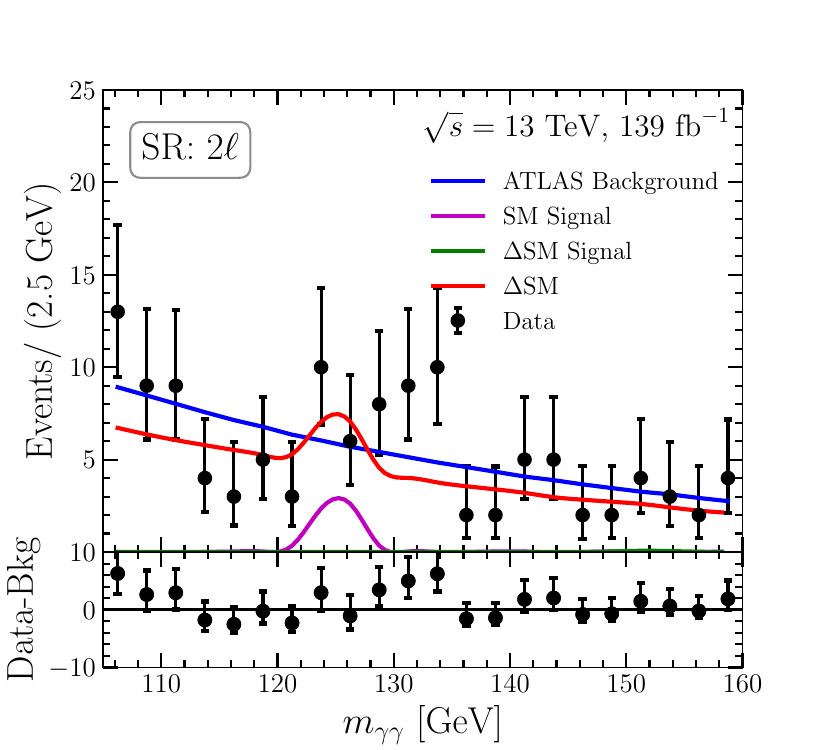}
\includegraphics[scale = 0.32]{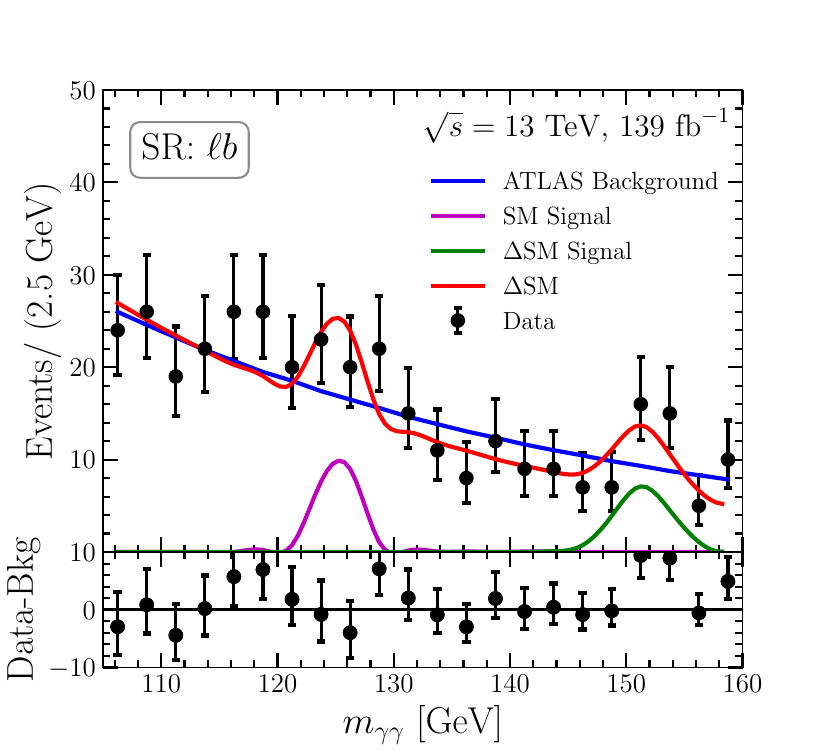}
\includegraphics[scale = 0.32]{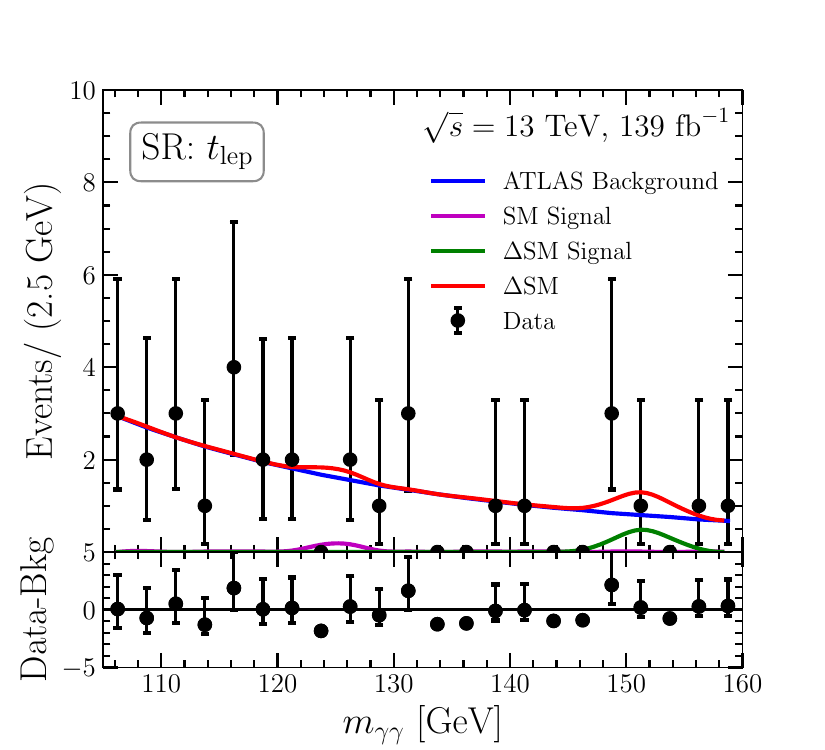}
\includegraphics[scale = 0.32]{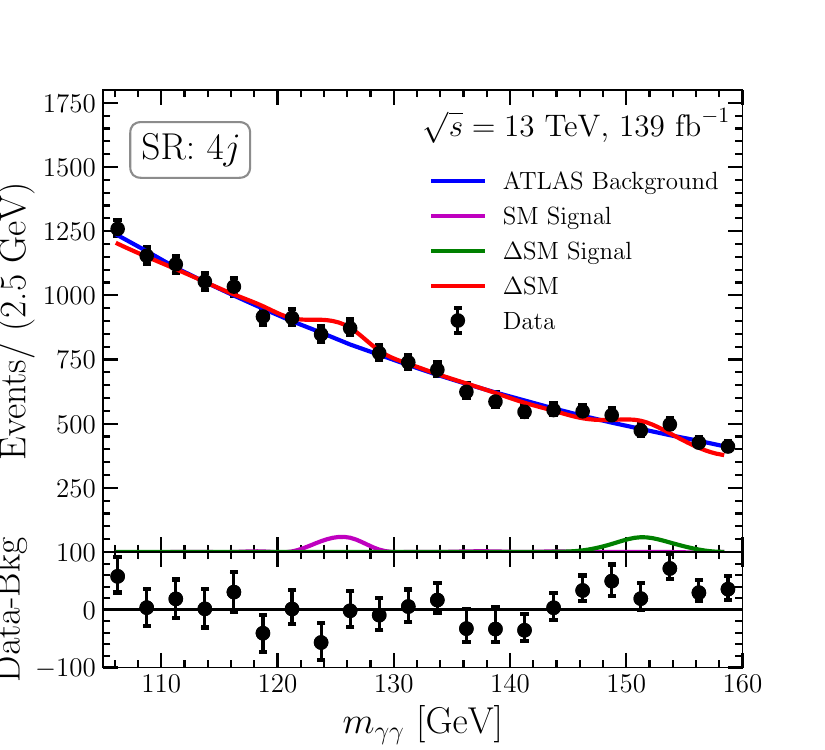}
\includegraphics[scale = 0.32]{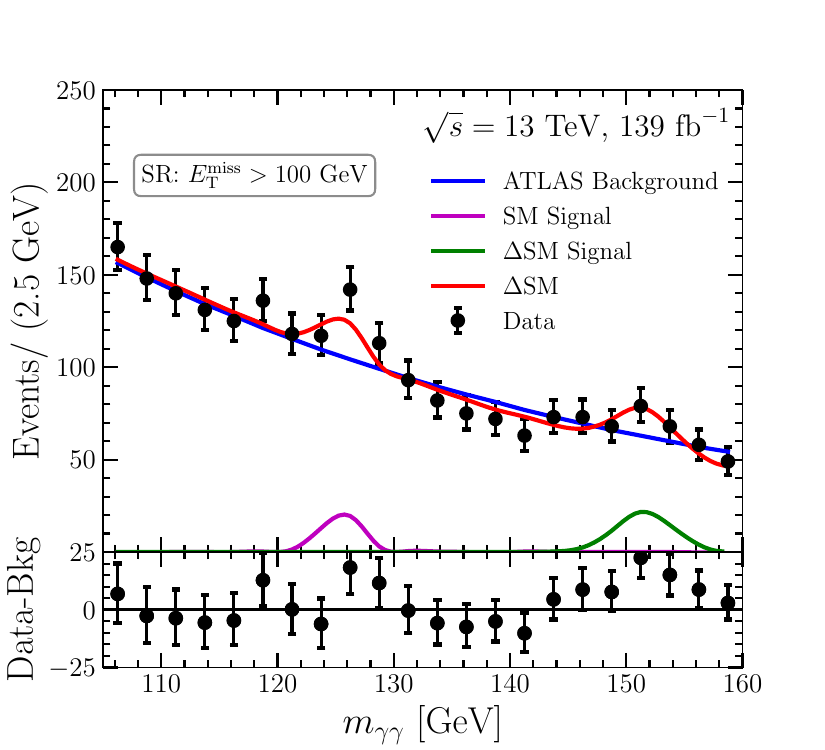}
\includegraphics[scale = 0.32]{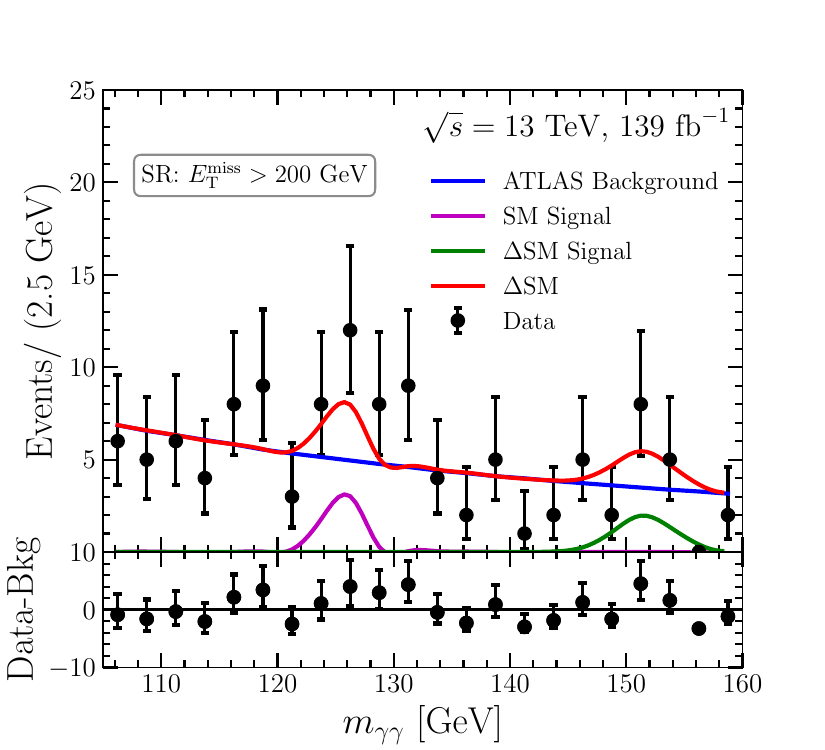}
\caption{Di-photon invariant mass distributions for eight relevant signal regions. The data (black) is shown together with the continuum background (blue) from the ATLAS analyses and the total $\Delta$SM events (red). The latter is comprised of the refitted background (not shown for brevity), the predicted SM Higgs signals at 125\,GeV (magenta) and 152\,GeV signal (green).}
\label{fig:fitted_eventdistribution}
\end{figure}

We consider searches for the production of photon pairs (di-photons) at the LHC in association with additional particles or missing energy.\footnote{Multi-lepton final states originating from the decays of the triplet Higgses to $WW,ZW$ and $tb$ were studied in detail in Ref.~\cite{Butterworth:2023rnw} finding that they can only exclude masses between $\approx160\,$GeV and $\approx200\,$GeV. In fact, Ref.~\cite{Butterworth:2023rnw} used $m_{\Delta^0}=m_{\Delta^\pm}=150\,$GeV as a benchmark point.} Since we are interested in the on-shell production of a new Higgs that decays to photons, we can look for a peak in the invariant mass spectrum of the di-photons. In this context, an extensive analysis of the associated production of the SM Higgs was performed in Ref.~\cite{ATLAS:2023omk}, containing 22 different channels ($\gamma\gamma+X$ where $X$ stands for leptons, missing energy, jets, {\it etc.}). In addition, recently, ATLAS released another search targeting various channels, including $\gamma\gamma+\tau$~\cite{ATLAS-CONF-2024-005}, which was not included in Ref.~\cite{ATLAS:2023omk}.\footnote{Ref.~\cite{ATLAS:2023omk} was done in the context of a non-resonant di-Higgs analysis and used a boosted decision tree (BDT) for categorizing events. To recast this analysis, we took the conservative approach to add the events of all three BDT cuts to recover the data set obtained by applying the event selection cuts.} The figures given in the ATLAS papers \cite{ATLAS:2023omk,ATLAS:2023omk} show the observed and expected number of events as a function of the invariant mass of the di-photon system between $105\,$GeV and $160\,$GeV, therefore covering our region of interest motivated by the multi-lepton anomalies.\footnote{We solely rely on ATLAS results here, because, unfortunately, no competitive CMS analysis for the associated production of a Higgs within our mass range of interest exists.}

\begin{figure}[htb!]
    \centering
    \begin{subfigure}{0.49\textwidth}{\includegraphics[width=\textwidth]{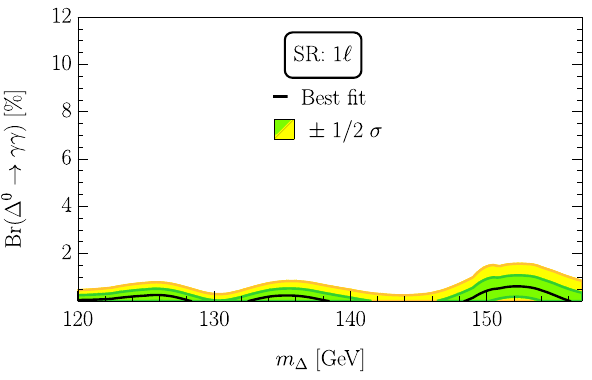}}\end{subfigure}
    \begin{subfigure}{0.49\textwidth}{\includegraphics[width=\textwidth]{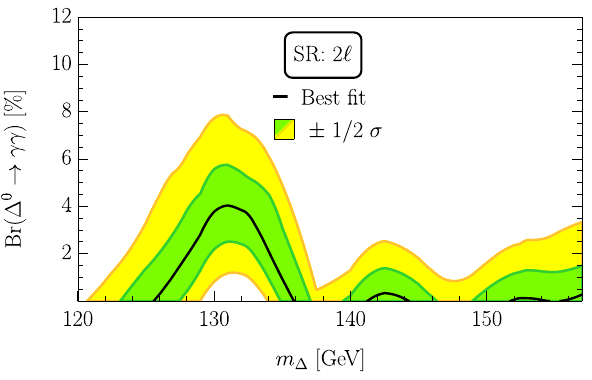}}\end{subfigure}
    \begin{subfigure}{0.49\textwidth}{\includegraphics[width=\textwidth]{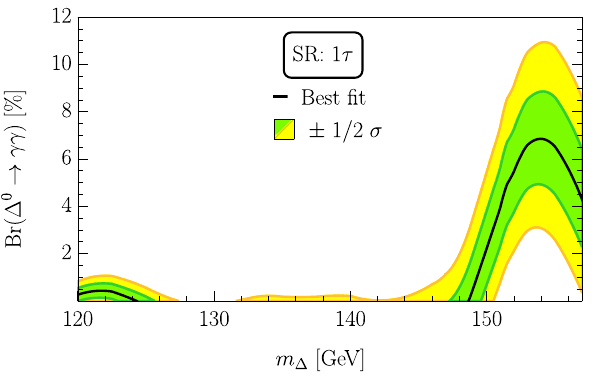}}\end{subfigure}
    \begin{subfigure}{0.49\textwidth}{\includegraphics[width=\textwidth]{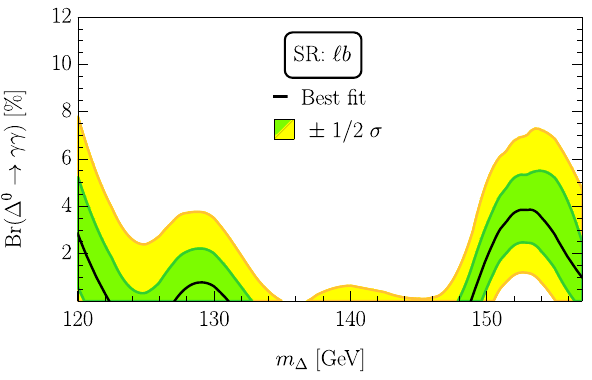}}\end{subfigure}
    \begin{subfigure}{0.49\textwidth}{\includegraphics[width=\textwidth]{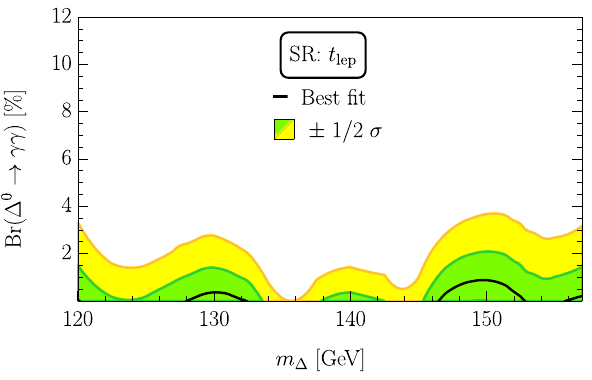}}\end{subfigure}
    \begin{subfigure}{0.49\textwidth}{\includegraphics[width=\textwidth]{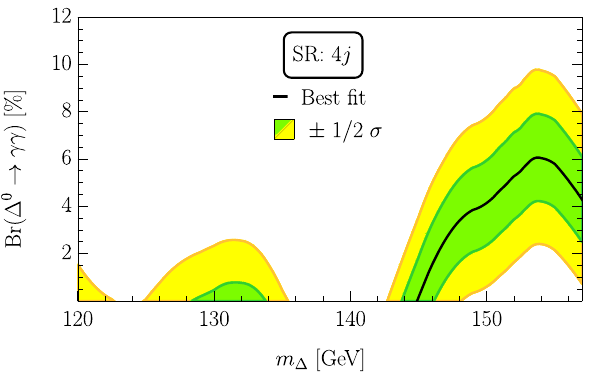}}\end{subfigure}
    \begin{subfigure}{0.49\textwidth}{\includegraphics[width=\textwidth]{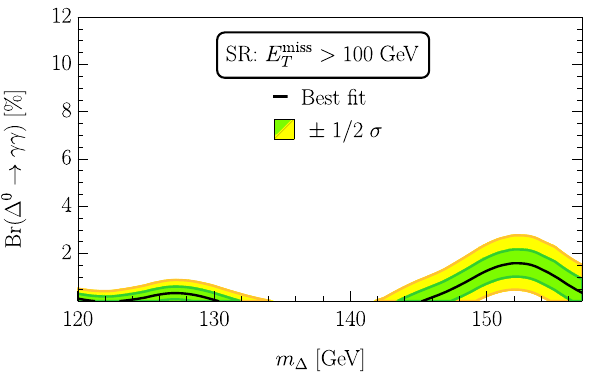}}\end{subfigure}
    \begin{subfigure}{0.49\textwidth}{\includegraphics[width=\textwidth]{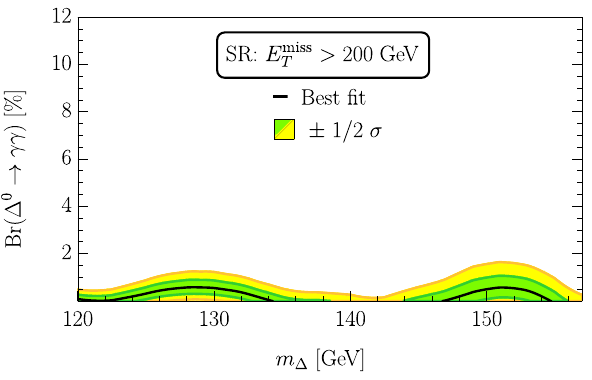}}\end{subfigure}
    \caption{Preferred range for the branching ratio of $\Delta^0 \to \gamma \gamma$ for the 8 signal regions which are sensitive to the signal of our model.}
    \label{fig:SRs}
\end{figure}

We simulated the processes $pp\to W^*\to (\Delta^\pm\to XY)(\Delta^0\to \gamma\gamma)$, with $\Delta^\pm$ decaying according to its (mass-dependent) branching ratios (see Fig.~\ref{fig:xsec}) using {\tt MadGraph5aMC@NLO}~\cite{Alwall:2014hca} with the parton showering performed by {\tt Pythia8.3}~\cite{Sjostrand:2014zea} and carried out the simulation for the ATLAS detector~\cite{ATLAS:2023omk} with {\tt Delphes}~\cite{deFavereau:2013fsa}. The UFO model file at NLO for the $\Delta$SM was built using {\tt FeynRules}~\cite{Degrande:2011ua,Alloul:2013bka, Degrande:2014vpa}.\footnote{To reduce the computational resources needed for our scan, we performed the simulations at leading order but rescaled the production cross-section to account for the NLO and NNLL effect following Refs.~\cite{Ruiz:2015zca,AH:2023hft}. Furthermore, we simulated the 4-jet category at NLO, where gluon radiation is particularly important, and included it via a correction factor of 1.2.} Taking into account that $Z$ and $W$ bosons decay (according to their known branching ratios) to leptons, missing energy and jets, we expect that the ATLAS signal regions targeting leptons, missing transverse energy ($E_T^{\rm miss}$) and high jet activity are the most sensitives ones. Furthermore, at higher values of $m_{\Delta}$, the categories addressing top quarks become relevant.\footnote{We did not include the signal region targeting hadronically decaying top quarks in our analysis. Here, ATLAS uses a BDT which targets top-pair production with a tight cut on the BDT score of 0.9. Because our signal consisting of a bottom quark and an off-shell top is quite different from this, the resulting efficiency is expected to be very small. Furthermore, we also used the single lepton category from Ref.~\cite{ATLAS-CONF-2024-005}, rather than from Ref.~\cite{ATLAS:2023omk}, since the bottom-quark jet veto leads to a nearly uncorrelated data set w.r.t.~the $\ell b$ category of Ref.~\cite{ATLAS:2023omk}.} In fact, we find that, of the 23 categories, the 8 listed in Table~\ref{categories} turn out to be relevant in our model for the mass range under consideration. The di-photon invariant mass distributions for these relevant signal regions are shown in Fig.~\ref{fig:fitted_eventdistribution}.

The backgrounds (including the SM Higgs for our purpose) given by ATLAS were obtained under the hypothesis that there is only a single resonance at 125\,GeV. Since we assume, in addition, a second resonance with a different mass and signal strength, the background has to be redetermined. For this, we subtract from the measured number of events per bin ($N_i^{\rm exp}$) the predicted number of SM Higgs events as well as the new physics signal (depending on $m_\Delta$ and Br[$\Delta^0\to \gamma\gamma$]) and fit this the continuous function
\begin{equation}
    \left(1 - \frac{m}{s}\right)^{b} + (m/s)^{a_0 +  a_1\log (m/s)}\,\label{background}
\end{equation}
with the free parameters $a_0, a_1$ and $b$, and $s=13$\,TeV being the LHC run-2 center-of-mass energy, and $m$ the invariant mass of the di-photon pair. In Fig.~\ref{fig:fitted_eventdistribution}, we show the fit to the di-photon invariant mass distributions for eight relevant signal regions. The continuum backgrounds taken from the ATLAS analyses are in blue, and the overall fitted $\Delta$SM signal-plus backgrounds are in red. Also shown are the 125 GeV SM Higgs and 152 GeV triplet Higgs signals.

To find (for a given mass) the preferred range for Br$(\Delta^0\to \gamma\gamma)$, we perform for each signal region a likelihood-ratio test using Poisson statistics. Thus, the theory prediction for the number of events in a bin $i$ (including the continuous background of Eq.~(\ref{background}), the SM Higgs signal and the new physics signal from the triplet Higgs) corresponds to the mean of the Poisson distribution, and we calculate the ratio
\begin{equation}
    \mathcal{L}_R=\Pi_i\left[\mathcal{L}(N_i^{\rm SM},N_i^{\rm exp})/\mathcal{L}(N_i^{\rm NP},N_i^{\rm exp})\right].
\end{equation}
Here $\mathcal{L}$ is the likelihood described by the Poisson distribution, $N_i^{\rm SM}$ ($N_i^{\rm NP}$) is the number of expected events in the SM ($\Delta$SM) and $N_i^{\rm exp}$ is the number of events measured by ATLAS. 

\begin{figure}[t!]
\centering
\includegraphics[scale=0.8]{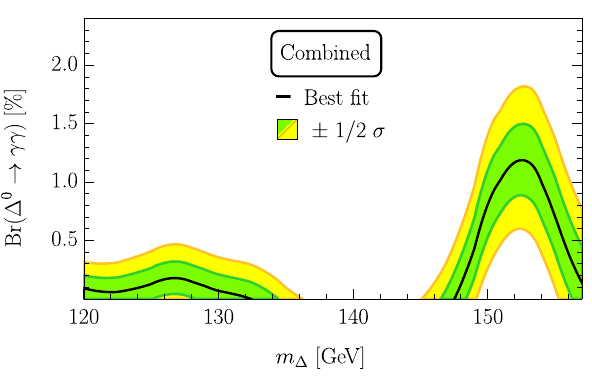}
\caption{Statistical combination of the 8 relevant channels including their correlations within the $\Delta$SM. Note that a non-zero branching ratio of $\Delta^0\to \gamma\gamma$ is preferred at $m_\Delta\approx152$\,GeV with a significance of $4.3\sigma$.}
\label{fig:exclusion_comb}
\end{figure}

The resulting 68\% and 95\% confidence level (CL) regions of Br$(\Delta^0\to \gamma\gamma)$, calculated by requiring that $\Delta\chi^2=-2\ln(\mathcal{L}_R)=1$ and $\Delta\chi^2=-2\ln(\mathcal{L}_R)=4$, respectively, are shown in Fig.~\ref{fig:SRs}.\footnote{Note that we allow for an unphysical negative branching ratio to take into account the effect of downward fluctuations of the background.} Interestingly, all relevant distributions display a preference for a non-zero decay rate to $\gamma\gamma$ at $\approx$152\,GeV. Combing all 8 signal regions, including the relevant correlations among them, we obtain the best-fit range for Br$(\Delta^0\to \gamma\gamma)$ as a function of $m_\Delta$ in Fig.~\ref{fig:exclusion_comb}. One can see a strong preference for a non-zero signal strength, which is most pronounced at $\approx$152\,GeV with a corresponding significance of $4.3\sigma$. 

\section{Discussion, Conclusions and Outlook}

The obtained $4.3\sigma$ for a new Higgs at $\approx$152\,GeV within the $\Delta$SM is the highest significance for a narrow resonance at the LHC  within a simple but ultraviolet-complete extension of the SM. Note that,  due to the mass range and the final state signatures predicted by the multi-lepton anomalies, the look-elsewhere effect is negligible. Consequently, we were able to combine the different channels without a penalty for additional degrees of freedom. Furthermore, the $\Delta$SM can account for the observed positive shift in $W$ and lead to a strong first-order phase transition required for generating a matter anti-matter asymmetry via Baryogenesis.

\begin{figure*}[t!]
\centering
\includegraphics[scale=0.45]{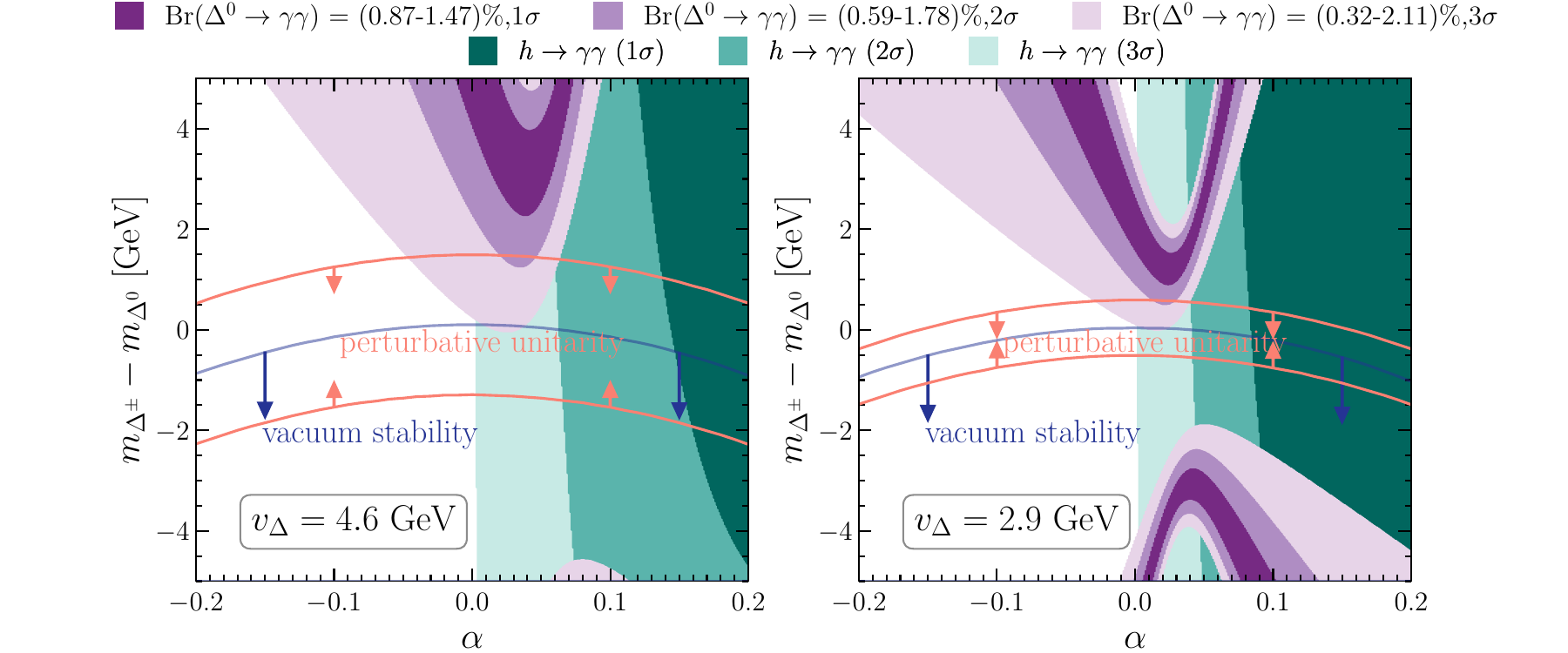}
\caption{\label{fig:H152Za} Preferred regions in the $\alpha$ vs $m_{\Delta^\pm}-m_{\Delta^0}$ plane for $m_{\Delta^0} = 152$ GeV and two values of $v_\Delta$: 4.6\,GeV (left) and 2.9\,GeV (right), corresponding to the central value obtained from the global electroweak fit with and without including the CDF-II measurement. The band between the two orange lines satisfies perturbative unitarity and the one below the blue line leads to a stable vacuum at the electroweak scale. The green regions are allowed by the SM Higgs signal strength to photons ($h \rightarrow \gamma \gamma$)~\cite{CMS:2021kom,ATLAS:2022tnm} signal strength at $1\sigma$ (1.02-1.15), $2\sigma$ (0.96-1.22) and $3\sigma$ (0.90-1.29) levels. The $1\sigma$, $2\sigma$ and $3\sigma$ regions for Br$(\Delta^0\to \gamma\gamma)$ are shown in violet. Note that both Br$(\Delta^0\to \gamma\gamma)$ and Br$(h \rightarrow \gamma \gamma)$ are sensitive to the mass splitting since this determines the tri-linear Higgs coupling which enters the processes via the $\Delta^\pm$ loop.}
\label{alpha-m}
\end{figure*}

So far, we have subsumed the free parameters of our model into Br$[\Delta^0\to\gamma\gamma]$. Let us now have a closer look at how the preferred decay rate for photons can be obtained. Br$[\Delta^0\to\gamma\gamma]$ is only a function of the mixing angle $\alpha$ and the mass splitting between the charged and neutral components of the triplet (for a given mass and $v_\Delta$). Looking at the corresponding parameter space for two representative values of $v_\Delta$ shown in Fig.~\ref{alpha-m}, one can see that the 2$\sigma$ region preferred by the associated di-photon production can lead to an unstable vacuum and only slightly overlaps within the non-perturbative regime. This indicates that while we have growing evidence for a 152\,GeV Higgs produced via Drell-Yan in association with leptons (e,$\mu$ and $\tau$), missing energy and $b$ quarks, the $\Delta$SM should be extended. In fact, while it can lead to a strong first-order phase transition, it cannot account for the additional charge-parity violation~\cite{Sakharov:1967dj} to obtain Baryogenesis since the triplet interactions are all real. This means that while the $\Delta$SM is not expected to be the final theory of Nature, also because it does not solve several of the problems of the SM mentioned in the introduction, it provides an important indication of the direction in which the SM should be extended.

A more comprehensive model addressing these problems could be the $\Delta$2HDMS model, where the SM is extended not only by a triplet but also by a singlet and a second doublet. This model can successfully give rise to weak-scale Baryogenesis~\cite{Inoue:2015pza} and explain the tensions between the theory predictions of the differential top-quark distribution and their measurements~\cite{Coloretti:2023yyq} which are part of the multi-lepton anomalies. Furthermore, the charged Higgs coming from the second $SU(2)_L$ doublet can modify Br$(\Delta^0\to \gamma\gamma)$ significantly (via the tri-linear term $H_1^\dagger\Delta^0H_2$) to avoid problems with vacuum stability and perturbative unitariy.

Finally, a new Higgs boson in general, and our $\Delta$SM in particular, significantly strengthens the physics case for future particle experiments. The charged component of the triplet could be examined with great precision at future $e^+e^-$ accelerators, such as the Circular Electron-Positron Collider (CEPC)~\cite{CEPCStudyGroup:2018ghi,An:2018dwb}, the Compact Linear Collider (CLIC)~\cite{CLICdp:2018cto}, the Future Circular Collider (FCC-ee)~\cite{FCC:2018evy,FCC:2018byv} and the International Linear Collider (ILC)~\cite{ILC:2013jhg,Adolphsen:2013jya}. Furthermore, these colliders can be used to determine the properties of the SM Higgs very precisely (which are, in general, altered in our model), produce many top quarks within a clean environment to test the related multi-lepton anomalies and clarify the preference of new physics in the global electroweak fit, in particular the $W$ mass where a positive-definite shift w.r.t.~the SM is predicted.

\bmhead{Acknowledgements}
The work of A.C.~is supported by a professorship grant from the Swiss National Science Foundation (No.\ PP00P21\_76884). A.C.~thanks Daniel Litim for useful discussion concerning vacuum stability. The work of S.A.~is partially supported by the National Natural Science Foundation of China under grant No.~11835013. S.P.M.~acknowledges using the SAMKHYA: High-Performance Computing Facility provided by the Institute of Physics, Bhubaneswar.

\bibliography{sn-bibliography}%
\end{document}